\documentclass[aps,prl,floatfix,twocolumn,superscriptaddress]{revtex4-2}
\usepackage{graphicx}
\usepackage{float}
\usepackage{amssymb,amsmath,stmaryrd,tabularx}
\usepackage[dvipsnames]{xcolor}
\usepackage{hyperref}

\makeatletter

\begin{document}
 \title{Remembrance of Tasks Past in Tunable Physical Networks}
\author{Purba Chatterjee}
\affiliation{Department of Physics and Astronomy, University of Pennsylvania, Philadelphia, PA $19104$, USA}

\author{Marcelo Guzman}
\affiliation{Department of Physics and Astronomy, University of Pennsylvania, Philadelphia, PA $19104$, USA}

\author{Andrea J. Liu}
\affiliation{Department of Physics and Astronomy, University of Pennsylvania, Philadelphia, PA $19104$, USA}
\affiliation{ Santa Fe Institute, 
Santa Fe, NM 87501, USA}

\begin{abstract}
Sequential learning in physical networks is hindered by catastrophic forgetting, where training a new task erases solutions to earlier ones.
We show that we can significantly enhance memory of previous tasks by introducing a hard threshold in the learning rule, allowing only edges with sufficiently large training signals to be altered.
Thresholding confines tuning to the spatial vicinity of inputs and outputs for each task, effectively partitioning the network into weakly overlapping functional regions. 
Using simulations of tunable resistor networks, we demonstrate that this strategy enables robust memory of multiple sequential tasks while reducing the number of edges and the overall tuning cost.
Our results hint at constrained training as a simple, local, and scalable mechanism to overcome catastrophic forgetting in tunable matter.
\end{abstract}

\maketitle
In physical networks, the interactions between nodes are dictated by the properties of the connecting edges. For example, nodes in a resistor network connected by resistors characterized by their conductances. As a whole, such networks can “compute” voltages at designated output nodes given voltages applied to input nodes by simply obeying their natural dynamics, driven by the physical imperative to satisfy Kirchhoff's laws. 

In {\it tunable} resistor networks, the conductances become ``tunable degrees of freedom," adjusted individually to obtain a desired output response to applied inputs---they are tuned to satisfy a \textit{task}~\cite{goodrich2015tuning,rocks2017designing,pashine2019directed}. 
``Self-tuning" networks have edges that can adjust their tunable degrees of freedom on their own, according to a local rule~\cite{stern2023learning}. Such networks can perform \textit{autonomous} supervised~\cite{pashine2019directed, pashine2020nonlinear,hexner2021periodic, dillavou2022demonstration,dillavou2024machine, kaiser2022hardware, niu2024self,cherry2025supervised} and unsupervised~\cite{guzman2025unsupervised} learning, thereby bypassing the need for central processors and memory storage and potentially leading to scalable self-tuning networks.

Multiple tasks can be trained in a single network, either simultaneously, by switching among them until a joint solution is found \cite{dillavou2025understanding, falk2023learning}, or sequentially, where tasks are learned one after the other~\cite{kortge2022episodic,french1991using,lewandowsky2014gradual,mcnaughton1995there,kumaran2016learning,robins1995catastrophic,kirkpatrick2017overcoming,serra2018overcoming,jedlicka2022contributions}. 
However, sequential training is often unsuccessful due to forgetting: learning a new task modifies the tunable degrees of freedom and can move the system away from the solution of previous tasks.
Forgetting can be abrupt, as in {\it catastrophic forgetting}~\cite{french1999catastrophic, laurier1continual}.
In artificial neural networks, strategies to address this problem rely on reducing representational overlap \cite{kortge2022episodic,french1991using,lewandowsky2014gradual}, or protecting weights that are important for previous tasks\cite{mcnaughton1995there,kumaran2016learning,robins1995catastrophic,kirkpatrick2017overcoming,serra2018overcoming,jedlicka2022contributions}. However, these approaches require global information and entail heavy memory storage and computational costs.

Here we focus on supervised learning and demonstrate a simple, powerful strategy to enhance memory of previously-trained tasks in tunable resistor networks.
We show that reducing the sensitivity of updates of tunable degrees of freedom---ignoring training signals below a threshold---enhances memory of previously learned tasks.
This principle exploits the nature of physical networks, where physical responses, and hence learning signals, decay with the distance from applied inputs. 
Intuitively, a threshold spatially confines training updates. As a consequence, multiple tasks can be learned sequentially by concentrating the tuned edges for each task in different spatial regions. 
Note that a threshold is a local constraint---each tunable resistor only needs to assess its own training signal---and does not rely on the specifics of the tuning process.

\textit{Systems studied---}
We consider ohmic (linear) resistor networks of $N$ voltage nodes and $N_E$ resistive edges with tunable conductances $\{k_1,k_2,...,k_{N_E}\}$ that are modified during training.
Nodes have fixed positions in space and edges connect nearby nodes.
Throughout the paper, we work with disordered planar networks with periodic boundary conditions.

A task is defined as a voltage input-output relation. We start with ``edge-coupling," where the input is defined as an imposed voltage drop $V_S$ across two adjacent source nodes, and the output is the voltage drop $V_T$ across adjacent target nodes at some distance from the source. The desired input-output relation is $V_T = \Delta V_S$, with coupling value $\Delta$~\cite{rocks2019limits,rocks2021hidden, rocks2017designing}.
For resistor networks, Kirchhoff's and Ohm's laws imply a linear response $V_T=\mathcal R V_S$, where $\mathcal R$ is a scalar (or matrix for multiple inputs/outputs) that depends nonlinearly on all conductances~\cite{stern2025physical,guzman2025microscopic}.

The performance of a task is quantified by the error between the physical output response and the desired output.
For edge-coupling, we consider the mean-square error, $\mathcal E=(\Delta V_S-\mathcal RV_S)^2/2$.
Training thus corresponds to updating the conductances such that the error is minimized.
In general, this is achieved by updates of the form $k_i\rightarrow k_i+\Delta k_i$, with  $\Delta k_i= -\gamma\partial C/\partial {k_i}$ being the training signal, $C$ a cost function sharing the same minimum as the error $\mathcal E$ (for gradient descent $C=\mathcal E$), and $\gamma$ the training rate.

In this work, we consider tuning rules of the type 
\begin{equation}\label{eq:learningrule}
\Delta k_i=
    \begin{cases}
    -\gamma\partial \mathcal{C}/\partial k_i,& \text{if } |\partial \mathcal{C}/\partial k_i|>\lambda\\
    0,              & \text{otherwise}
    \end{cases}.
\end{equation}
Thus, the conductance of an edge is updated only if magnitude of its gradient exceeds a fixed threshold $\lambda$. 
This criterion is fundamentally local, making it suitable for local tuning rules~\cite{stern2021supervised, guzman2025unsupervised, scellier2017equilibrium, pashine2019directed}, although thresholds can also be introduced into global tuning processes.

In all reported simulations we choose $C$ to be the contrastive power~\cite{scellier17,stern2021supervised},  leading to the Coupled Learning rule~\cite{stern2021supervised} implemented experimentally in Contrastive Local Learning Networks~\cite{dillavou2024machine}.
Details of the training protocol are given in the SI.

\begin{figure}
\begin{center}
\includegraphics[scale=0.45]{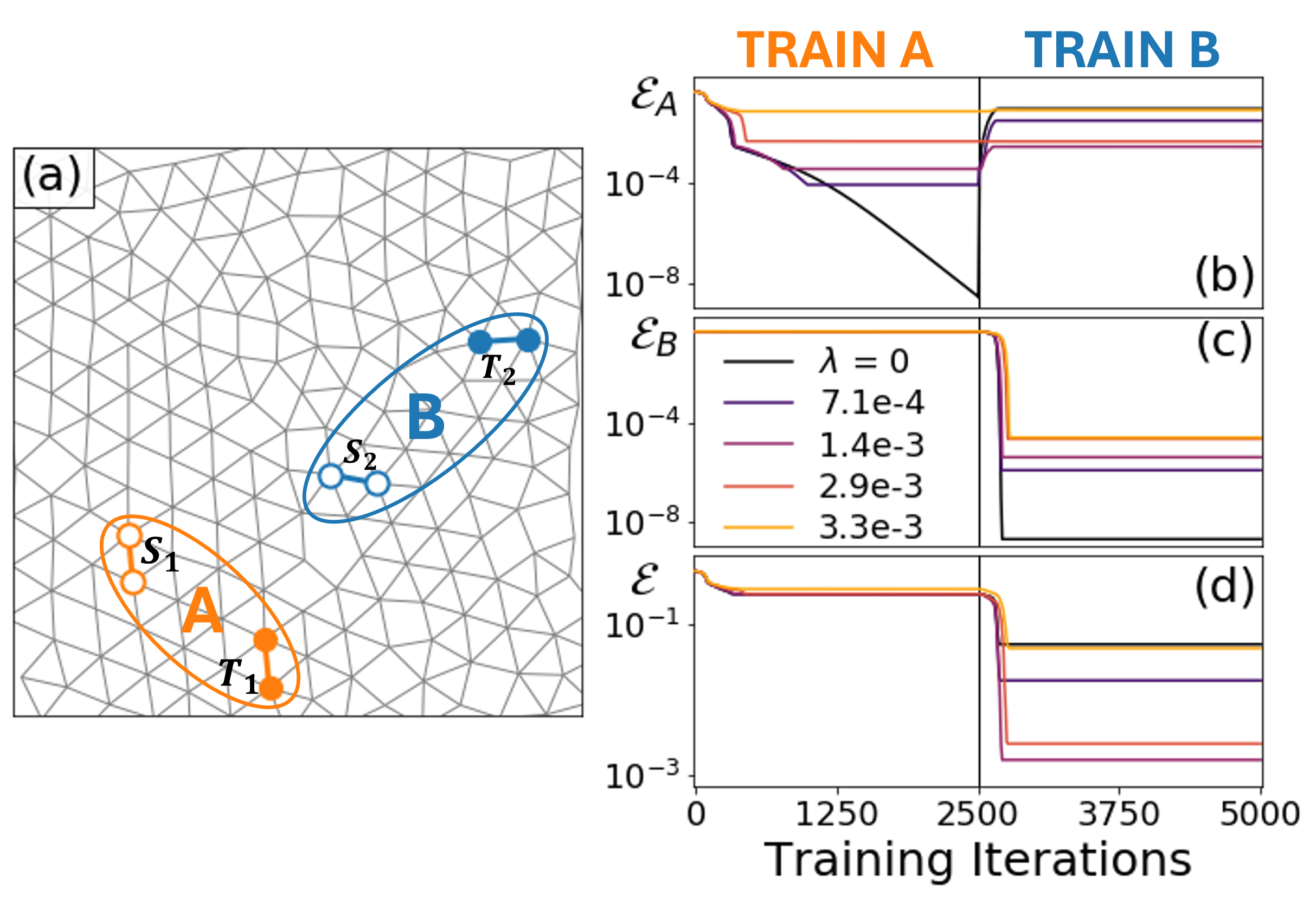}
\end{center}
\caption {
{\bf Sequential learning of two edge-coupling tasks.}
(a) Network of $N=256$ nodes and  $N_E=704$  edges, trained for two sequential edge-coupling tasks $A$ (orange) and $B$ (blue). Source nodes are hollow circles and target nodes are solid circles. (b-d) From top to bottom: Errors $\mathcal{E}_A$, $\mathcal{E}_B$, and joint error $\mathcal{E}$ as a function of training steps $T$ for different thresholds $\lambda$ (color).
The lowest final joint error is found for $\lambda =0.0014$.}
    \label{fig:twotasks_specific}
\end{figure}

\textit{Thresholding in single and multiple tasks---}
We begin with a network trained sequentially for two edge-coupling tasks $A$ and $B$, with coupling value $\Delta=1$ for both, see Fig.~\ref{fig:twotasks_specific}a. 
We first train the network to minimize the error $\mathcal E_A$ and then $\mathcal E_B$. The overall performance is defined as the joint error, $\mathcal E=(\mathcal E_A+\mathcal E_B)/2$.

The effects of the threshold are shown in Fig.~\ref{fig:twotasks_specific}b,c,d, for several values of $\lambda$.
In the absence of a threshold ($\lambda =0$) the network has maximal sensitivity, adapts to the smallest gradients, and therefore reaches the lowest error $\mathcal E _A$  for task A.
However, subsequent training on task B drives $\mathcal E_A$ back to values comparable to the untrained case, while lowering $\mathcal E_B$.
The network learns both tasks well in isolation but almost completely forgets task A when trained for B, leading to large joint error $\mathcal E$.
In contrast, for $\lambda>0$ there is a tradeoff: the network cannot reach the lowest individual task errors, but the joint error can be lowered by up to two orders of magnitude relative to the unconstrained case. The final joint error $\mathcal E$ is non-monotonic in $\lambda$ and there is a range of thresholds around $\lambda = 1.4 \times 10^{-3}$ that optimally allows task A to be remembered after training task B.

\begin{figure}
\begin{center}
\includegraphics[scale=0.45]{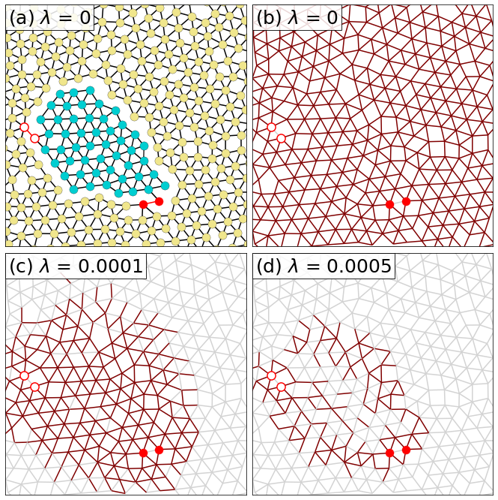}
\end{center}
\caption{{\bf Increasing the threshold localizes effects of training.}
(a) Network with $N=1024$ nodes and $N_E=2824$ edges trained for an edge-coupling task. The width of each edge is proportional to its final conductance. A low-conductance boundary separates sectors of positive (blue nodes, $V=0.5$) and negative  (yellow nodes, $V=-0.5$) voltages. 
(b-c) Snapshots of the network at the end of training for different thresholds.  Edges that changed their conductance during training are shown in maroon. For $\lambda=0$ every edge changes. The number of altered edges decreases with increasing threshold $\lambda$.}
\label{fig:singletask_lam}
\end{figure}
To understand this result it is instructive to analyze the effect of thresholding on a single edge-coupling task.
Fig.~\ref{fig:singletask_lam}a shows the physical response of a trained network in the absence of any threshold ($\lambda=0$) for an edge-coupling task with $\Delta =1$ and homogeneous initial conductances, $k_i=1,\;\;\forall i=1,...,N_E$.
The tuned structure is partitioned into two sectors with nearly uniform node voltages, separated by a low-conductance crack given by the minimum allowed value $k=10^{-6}$, as reported in~\cite{rocks2021hidden}.
With one source and target node in each sector, the voltage drop across the target and source nodes are nearly identical as required by training.
To achieve this configuration, all conductances in the network were changed during training, see Fig.~\ref{fig:singletask_lam}b.

For the same training at different thresholds $\lambda$,  Fig.~\ref{fig:singletask_lam}c and d show that the altered edges decrease in number and concentrate near the low-conductance boundary as $\lambda$ increases.
This reflects the locality of physical networks: training signals decay with distance from sources and targets, and distant edges have less influence on the task.
Thresholding therefore localizes changes to the region relevant for the task.
Note that this result is robust to the value of $\Delta$ (see SI).

\begin{figure}
\begin{center}
\includegraphics[scale=0.45]{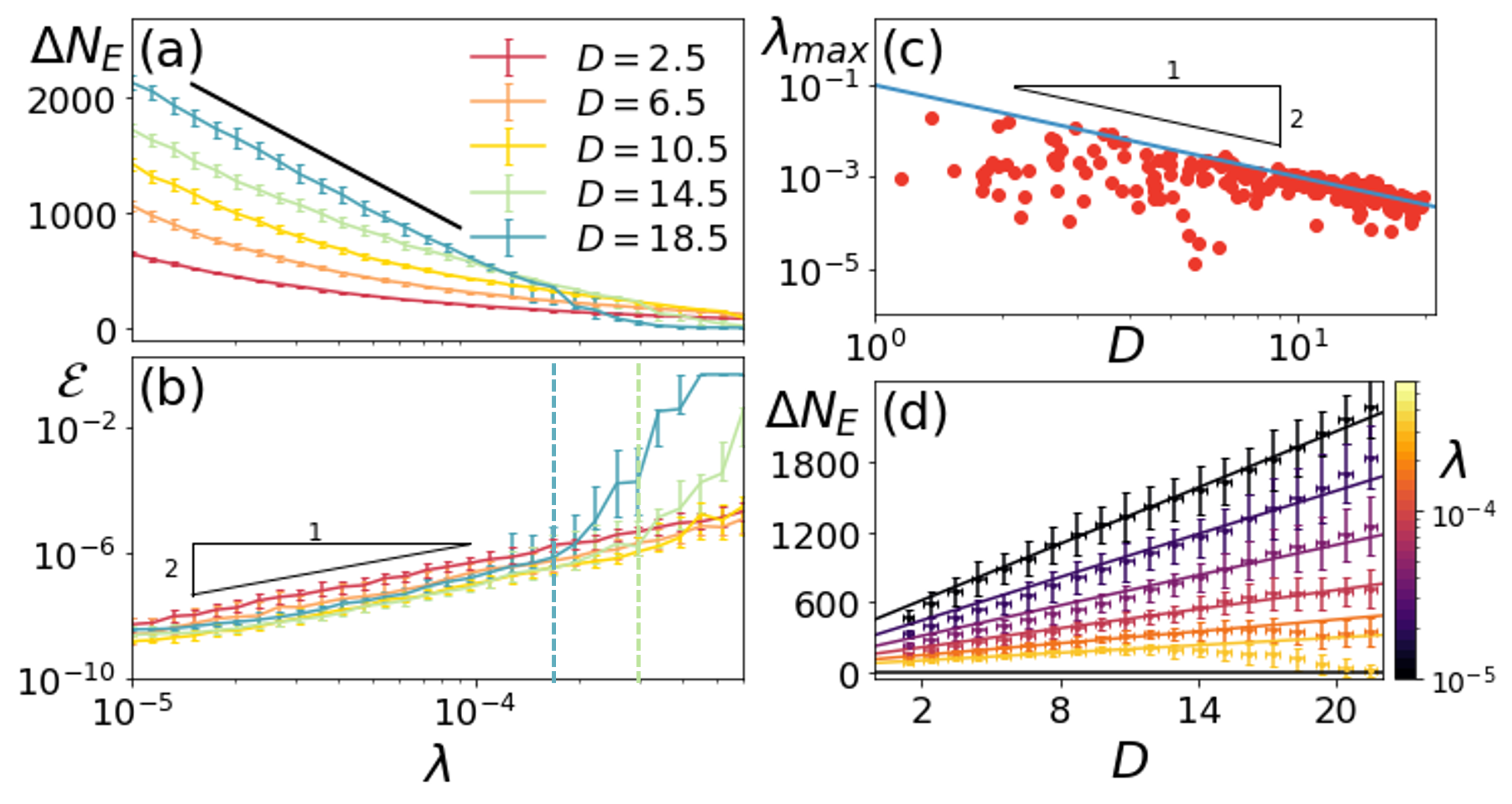}
\end{center}
\caption{{\bf Effects of threshold on a single edge-coupling task.}
Median values for the total number of altered edges $\Delta N_E$  (a) and error $\mathcal E$  (b)  as a function of $\lambda$, for a network of $N=1024$ nodes and $N_E=2824$ edges.
Color encodes the distance $D$ between source and target nodes.
Training fails for thresholds higher than a maximum value $\lambda_{max}$, indicated by dashed lines in (b) for $D=14.5$ (blue) and $D=18.5$ (green). 
(c) Maximum threshold $\lambda_{max}$ versus distance $D$. Each dot corresponds to a trained network.
(d) $\Delta N_E$ versus $D$ for different $\lambda$, as indicated by color. Error bars in (a,b,d) show first and third quartiles over $100$ realizations with randomly selected source-target pairs.
}
    \label{fig:singletask_D}
\end{figure}

Fig.~\ref{fig:singletask_D}a shows the total number of updated edges $\Delta N_E$ during training, as well as the final error $\mathcal E$, for varying values of $\lambda$ and distances $D$ between  sources and targets. Here $D$ is the shortest path length between any pair of source-target nodes, reported in units of the average edge length in the network.
In general, thresholds larger than a distance-dependent critical value, $\lambda>\lambda_{max}(D)$, forbid updates of the key edges belonging to the low-conductance boundary.
For thresholds smaller than this critical threshold, we observed two opposite trends:  $\Delta N_E\sim -\log\lambda$ and $\mathcal E\sim \lambda^2$.
The value $\lambda_{max}(D)$ signals the deviation of the error from the quadratic scaling, see dashed lines in Fig.~\ref{fig:singletask_D}b.
This value decreases with distance as $\lambda_{max}(D)\sim D^{-2}$, see Fig.~\ref{fig:singletask_D}c.
In other words, because training signals decay with distance from the applied inputs, more sensitivity (smaller $\lambda$) is needed to train the desired task successfully at larger $D$.
For $\lambda<\lambda_{max}(D)$, $\Delta N_E$ is roughly linear in $D$ (Fig.~\ref{fig:singletask_D}d).

\begin{figure}
\begin{center}
\includegraphics[scale=0.45]{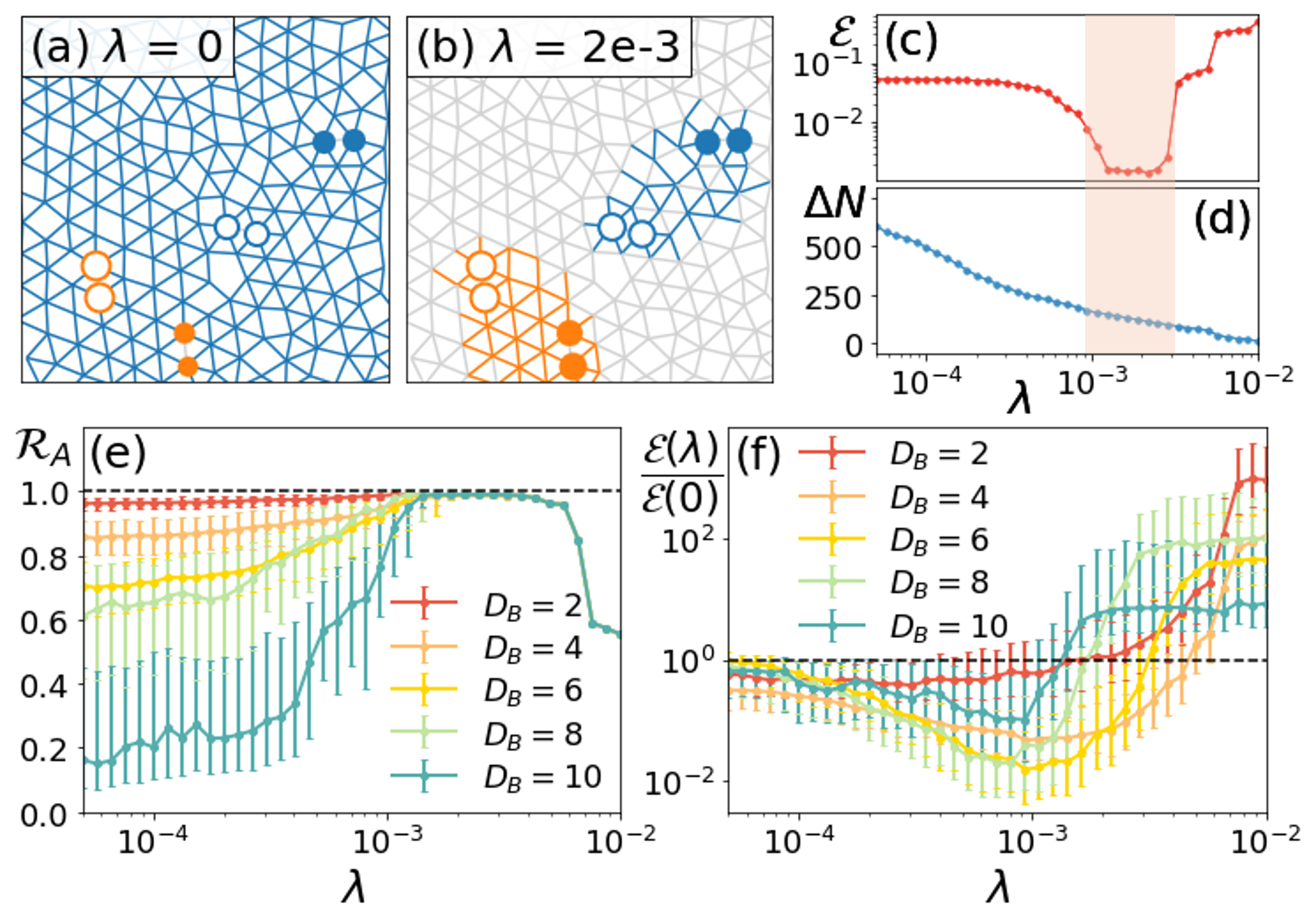}
\end{center}
\caption {A closer look at the tasks shown in Fig. ~\ref{fig:twotasks_specific}. 
(a,b) Final networks after sequential training of tasks A and B for $\lambda=0$ and $\lambda = 0.002$. Edges are colored blue if altered during the training of B, orange if altered only during A, and gray if unchanged.
(c,d) Joint error $\mathcal{E}$ and the total number of altered edges $\Delta N_E$ versus $\lambda$. 
For an intermediate range of $\lambda$ (shaded) the joint error is strongly reduced while adjusting only $15-20\%$ of the edges. 
(e) Median voltage response $\mathcal R_A$ (desired value $\Delta =1$) and (f) joint error $\mathcal E$ after full sequential training as a function of $\lambda$ for different distances $D_B$ (color) and fixed $D_A=4$.
Error bars show first and third quartiles over $100$ realizations with randomly selected sources and targets for task $B$ and fixed task $A$.}
    \label{fig:twotasks_specific_solutions}
\end{figure}

The single-task behavior explains the sequential learning performance shown in Fig.~\ref{fig:twotasks_specific}. For that specific choice of tasks A and B,
Figs.~\ref{fig:twotasks_specific_solutions}a and b show the edges that changed during training for $\lambda=0$ and $\lambda = 0.002$, respectively. 
At $\lambda=0$, all edges are modified for both tasks.
At $\lambda=0.002$, key edges for task A (orange) are unaltered when the system is trained for B, and the altered edges for each task largely occupy distinct regions.
More generally, joint error is significantly reduced over a range of values of $\lambda$, with merely $15-20\%$ of edges sufficient for optimal performance and memory, see Figs.~\ref{fig:twotasks_specific_solutions}c and d.

How much the memory of task $A$ is enhanced depends on the distance between sources and targets. 
Fig.~\ref{fig:twotasks_specific_solutions}e illustrates the physical response of task $A$, $\mathcal R_A$, as a function of $\lambda$ and varying distances $D_B$ between source and target of task $B$. 
Fig.~\ref{fig:twotasks_specific_solutions}e shows that 
$\mathcal R_A$ is close to the desired value $\Delta=1$ over the same range of thresholds for all $D_B$, but the improvement relative to $\lambda = 0$ case is larger at larger $D_B$.
In all cases, introducing a threshold improves the joint error regardless of $D_B$, see Fig.~\ref{fig:twotasks_specific_solutions}f.

\begin{figure}
\begin{center}
\includegraphics[scale=0.45]{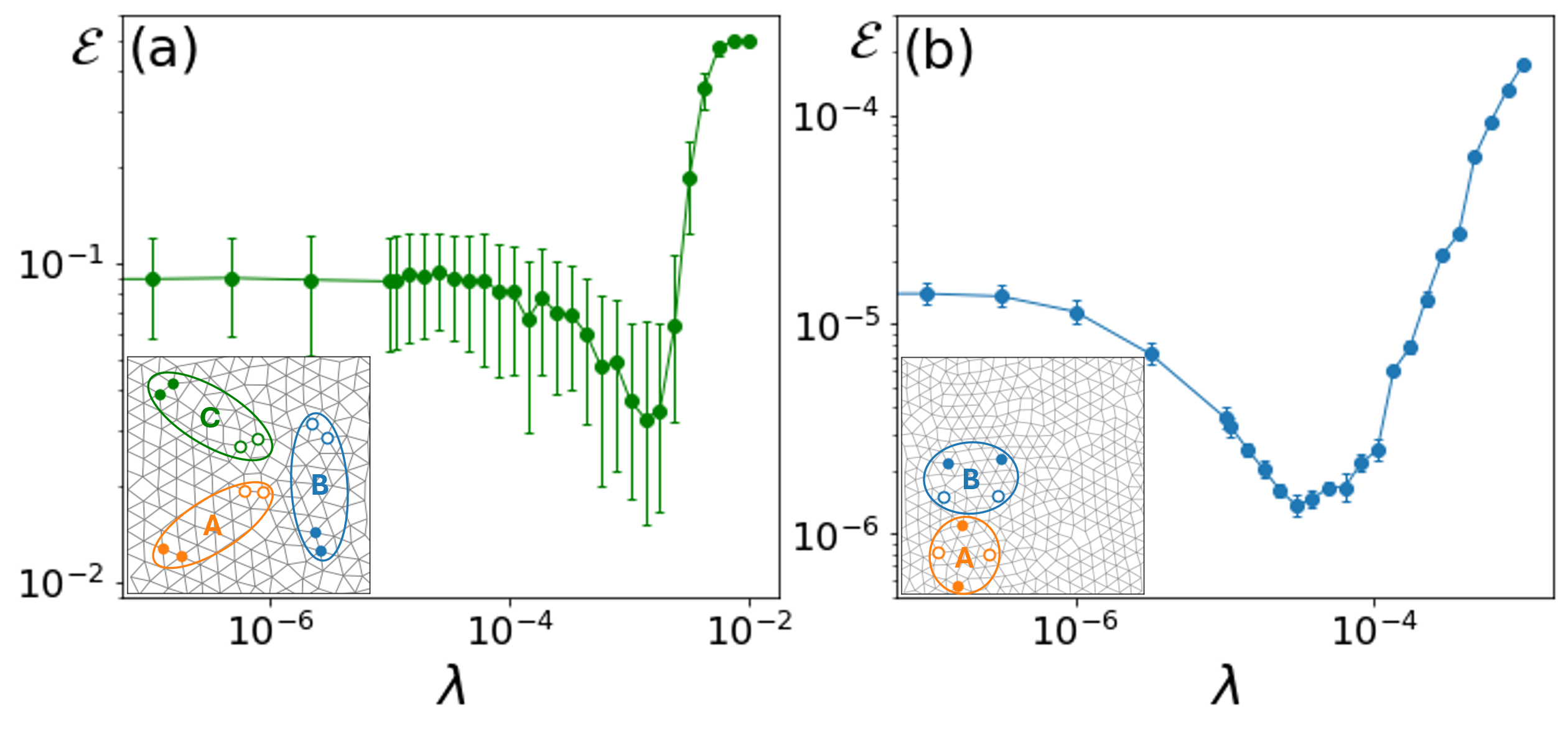}
\end{center}
\caption{ 
(a) Median joint error $\mathcal{E}$ as a function of $\lambda$ after training for three sequential edge coupling tasks $A$, $B$, and $C$ in a network with $N=256$ nodes and $N_E=704$ edges (inset).
The joint error, $\mathcal E=(\mathcal E_A+\mathcal E_B+\mathcal E_C)/3$, reaches a minimum at $\lambda=0.002$. Error bars represent the first and third quartiles computed from $100$ realizations with randomly chosen source-target pairs for each task.
(b) Median joint error $\mathcal{E}$ as a function of $\lambda$ after training sequentially for two linear regression tasks $A$ and $B$ (inset) in a network with $N=512$ nodes and $N_E=1405$ edges. The lowest error is found at $\lambda = 0.0001$.
Error bars represent the first and third quartiles computed from $100$ realizations, each with initial conductances drawn from a normal distribution with mean $1$ and standard deviation $0.05$.
}
    \label{fig:complextasks}
\end{figure}

The previous results extend beyond the sequential learning of two edge-coupling tasks.
In Fig.~\ref{fig:complextasks}a, we illustrate the joint error for three sequential edge-couplings ($\Delta =1$) as a function of $\lambda$.
Optimal performance is obtained for $\lambda \approx 0.002$. Above this value, $\mathcal{E}$ increases due to failure in the training of one or more tasks.

We also consider sequential training of more standard machine-learning tasks, linear regression.
Unlike edge-coupling, where the task is a fixed input-output relation, in linear regression the network learns a family of relations. 
For example, in a two-source two-target linear regression, we are interested in the physical response represented as
\begin{equation}
    \begin{pmatrix}
        V_{T_1}\\V_{T_2}
    \end{pmatrix}=\mathcal R \begin{pmatrix}
        V_{S_1}\\V_{S_2}
    \end{pmatrix},
\end{equation}
where $\mathcal R$ is now a matrix whose entries are functions of all the conductances of the network.
As before, the goal of linear regression is to learn a specific form of $\mathcal R=\Delta$ through samples.
We consider two sequential linear regression tasks on a network of $N=512$ nodes (see inset of Fig.~\ref{fig:complextasks}b), for which we define the desired relation as
\begin{equation}
    \Delta =\begin{pmatrix}
        0.1 & 0.2\\
        0.3 & 0.15
    \end{pmatrix},
\end{equation}
for both tasks, with different pairs of input and output nodes for the two tasks. 
Fig.~\ref{fig:complextasks}b shows that introducing a threshold improves the joint training error over many initial conditions by an order of magnitude.

\textit{Tradeoffs between memory, training time, and energy---}
We have already seen in Fig.~\ref{fig:singletask_D} that with thresholding, the significant decrease in the number of edges used to learn a task comes at the cost of larger errors. We now investigate how thresholding modifies two other important metrics in tunable networks: the number of learning iterations to achieve a given performance and the total energy required to apply these updates~\cite{stern2024training,dillavou2024machine}.

\begin{figure}
\begin{center}
\includegraphics[scale=0.43]{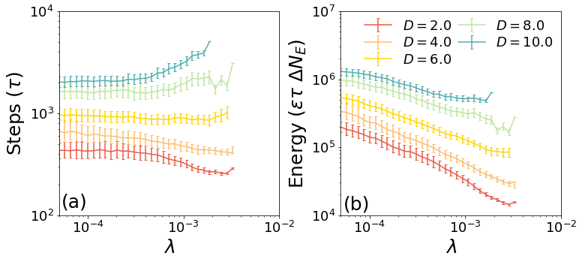}
\end{center}
\caption {{\bf Increasing the threshold reduces training energy.} Number of training steps $\tau$  (a) and an upper bound on the training energy $\epsilon \tau \Delta N_E$  (b) as a function of the threshold $\lambda$, for the successfully trained networks of Fig.~\ref{fig:twotasks_specific_solutions}. Colors indicate distances $D$ between source and target nodes. While $\tau$ remains approximately constant, the training energy decreases with increasing $\lambda$ for all values of $D$. Error bars represent the first and third quartiles computed from a subset of $100$ realizations with randomly selected source-target nodes for task $B$, for which training was successful for both tasks.
}
    \label{fig:tradeoffs}
\end{figure}

We focus on each edge-coupling example of Fig.~\ref{fig:twotasks_specific_solutions} and consider only networks and values of $\lambda$ that reduce the error per task at least to $10^{-4}$ of its initial untrained value.
We define $\tau_{A}$ as the number of training steps until all parameter updates for task $A$ stop.
By definition, $\tau_A\leq T$, where $T=10000$ is the maximum number of training iterations allowed per task. 
Analogously, we define $\tau_B$, as well as the total number of iterations to train the two tasks sequentially, $\tau =\tau_A+\tau_B$.
Fig.~\ref{fig:tradeoffs}a shows that $\tau$ remains approximately constant with $\lambda$, with slight increases for large values of $D$. Thus, improving memory through thresholding does not much affect training time. 

To estimate the total update energy cost, we assume that each edge update has a fixed energy cost $\epsilon$, independent of system size and magnitude of training signal.
The total energy cost of training is then given by $\sum_{t=0}^\tau\epsilon \delta N_E(t)$, where $\delta N_E(t)$ corresponds to the number of altered edges between training time $t$ and $t+1$.
Notice that this energy has an upper bound of $\epsilon\tau\Delta N_E$.
Fig.~\ref{fig:tradeoffs}b shows this upper bound as a function of $\lambda$, demonstrating that even when $\tau$ slightly increases with $\lambda$, the upper bound on the energy decreases. 
The dominant factor is the reduction of the number of altered edges with increasing $\lambda$. 
Introducing a threshold not only can improve memory in sequential learning, but  also reduces the energy cost of training in tunable networks. 

\textit{Discussion---} 
We have shown that reducing the sensitivity of updates in parameters to training signals can improve sequential learning in physical networks.
This result builds upon the local interactions of such networks and on the fact that physical responses and tuning signals decay with distance from inputs.
Through examples of edge-coupling and linear regression tasks, we have illustrated that ignoring weak training signals allows for a spatial modularity of the trained tasks.
There are two prerequisites to observe such modularity: First, the tasks have to take place in different locations, and second, the network needs to be sufficiently over-parameterized.
Indeed, we have shown that imposing a threshold on a tuning rule in a physical system effectively decreases the number of tunable edges for a given task---too few of them and the individual task cannot be learned.
Finally, in addition to improving the remembrance of tasks past, imposing a minimum threshold on parameter updates is advantageous from an energetic point of view.
With respect to the unconstrained case, our results show that raising the threshold may lead to more steps required for successful training, but reduces the number of updated edges.
The latter effect is dominant, so that the total number of edge updates during training {\it decreases} with threshold $\lambda$, reducing the energy required for training.

We have analyzed linear networks, but the considerations underlying the effect of thresholding should not be limited to the linear case. Generalizing our results to nonlinear resistor networks is an important direction for future work.

Our strategy is fundamentally different from a previous approach to obtain low-power solutions in resistor networks by introducing a tradeoff between task error and the power required to physically ``compute" outputs given inputs~\cite{stern2024training}. The two approaches are compatible, and can therefore be combined. Our results indicate that introducing a threshold for updates can be a useful strategy for autonomous learning platforms such as Contrastive Local Learning Networks composed of self-adjusting resistors. 
Furthermore, it can help counteract the undesired effects of physical imperfections~\cite{dillavou2025understanding}.
For example, many physical learning platforms are plagued with noise and bias that perturb training signals and thereby affect task performance~\cite{kaiser2022hardware, dillavou2025understanding}.
Implementing a threshold can reduce those effects without requiring any modeling of the physical system.

Our results are physically intuitive and should be quite general as they are not restricted to specific tuning processes, network types or architectures. 
In addition, they open up new questions. Persistent homology analysis of resistor~\cite{rocks2021hidden} and mechanical networks~\cite{rocks2024topological} trained for edge-coupling tasks reveals sectors of nearly uniform node voltage or strain. Thresholding appears to accentuate this sector structure; the persistence of sectors might be tied quantitatively to the ability to remember past tasks. 
Furthermore, elastic and resistor networks undergo random constraint-satisfaction transitions that limit the number of simultaneous tasks they can perform~\cite{rocks2019limits}.
It remains to be seen how the location and nature of the transition depends on threshold. 
Edge-coupling has also been investigated in the context of epistasis in mechanical networks~\cite{bravi2020direct,alqatari2024epistatic}, in which the effect of a mutation (perturbation of a conductance) depends on the mutation history.
Here, the existence of a threshold can potentially modify the interdependence of mutations as well as structural correlations that emerge for the learned solutions~\cite{rouviere2023emergence, guzman2025microscopic,yan2017architecture}.
Finally, and more broadly, we can ask whether various biological networks might use this strategy.

\textit{Acknowledgments---} This work was supported by DOE Basic Energy Sciences through grant DE-SC0020963 (PC and MG) and the Simons Foundation through Investigator grant \#327939 to AJL.

\bibliography{biblio}
\pagebreak

\onecolumngrid
\appendix

\setcounter{equation}{0}
\setcounter{subsection}{0}
\setcounter{figure}{0}
\renewcommand{\theequation}{A\arabic{equation}}
\renewcommand{\thefigure}{A\arabic{figure}}
\renewcommand{\thesubsection}{A\arabic{subsection}}

\section*{Appendix}

\subsection{Coupled Learning Rule and Simulation Details}
All resistor networks were trained using a contrastive local learning rule known as coupled learning~\cite{stern2021supervised}.
In short, the learning signal is proportional to the difference of the squared voltage drops of two physical responses:
\begin{equation}
    \Delta k_i = \frac{\gamma}{\eta}\left[(\Delta V_{F})_i^2-(\Delta V_C)_i^2\right],
\end{equation}
where $\gamma$ is the learning rate, $\eta$ a nudge parameter, $(\Delta V_F)_i$ the voltage drop at edge $i$ of the physical response $V_F$ known as ``free" state, and $(\Delta V_C)_i$ is the same quantity for the ``clamped" state.
The free state $V_F$ is obtained by imposing the input voltages at the source nodes.
The clamped state is obtained by imposing the same inputs as before while additionally imposing the output values of target nodes to be closer to the desired ones.
Denoting $V_{F,a}$ the voltage value of target node $a$ in the free state, and $\overline V_a$ the desired value, then in the clamped state the output nodes are fixed to
\begin{equation}
    V_{C_a}=V_{F,a}+\eta(\overline V_a-V_{F,a}).
\end{equation}
In all the reported simulations we used $\gamma =1$ and $\eta=10^{-4}$.

The results reported in all figures except Fig. $5$b in the main text, were generated with homogeneous initial conductances $k_i=1,\;\;\forall i=1,...,N_E$. For numerical stability, we imposed a lower bound $k=10^{-6}$ for the conductance of each edge during training.

For all edge-coupling tasks, we impose voltages $+0.5$ and $-0.5$ on the two source nodes, for a voltage drop $V_S=1$. Figs. 1, 4, 5a and 6 correspond to a single network with $N=256$ nodes and $N_E=704$ resistive edges, trained sequentially for edge-coupling tasks with $\Delta=1$, and $10000$ learning iterations allowed for each task. In the two-task cases of Figs. 4e,f and 6, we consider $100$ realizations, each with the sources and targets of task $A$ fixed, and the sources and targets of task $B$ chosen randomly for a given value of the separation $D_B$. In Fig. 6, only those realizations that lead to successful training---defined as the final error falling below $10^{-4}$ of the initial untrained error---are chosen for statistics. In Fig. 5a, we show statistics for $100$ realizations with randomly chosen sources and targets for each of the three edge-coupling tasks $A$,$B$ and $C$.

Figs. 2 and 3 correspond to a network with $N=1024$ nodes and $N_E=2824$ resistive edges trained for a single edge-coupling task with $\Delta=1$ and $100000$ allowed learning iterations. The statistics in Fig. 4 are over $100$ realizations with randomly chosen sources and targets with a given separation $D$.

Finally in Fig. 5b, we consider a network with $N=512$ nodes and $N_E=1405$ resistive edges, trained for two sequential linear regression tasks, each with the desired input-output relationship given by Eq. 2 and 3. We report statistics for $100$ realizations, each with initial conductances drawn from a normal distribution with mean $1$ and standard deviation $0.05$. The position of the source and target nodes of each task are given in the inset, and are fixed over the different realizations.

For all figures that contain statistics, we report the median values, and the first and third quartiles of the data as error.

\begin{figure}
\begin{center}
\includegraphics[scale=0.75]{sector_delta_1.0.png}
\end{center}
\caption {Final Network with $N=1024$ nodes and $N_E=2824$  edges, after training for an edge-coupling task with $\Delta =1$.  Each node is depicted by a solid circle proportional in size to the absolute value of their final voltage. Each solid edge's width is proportional to its final conductance, with altered edges in black and unaltered edges in gray. The dotted edges with $k=10^{-6}$ compose the low-conductance boundary that completely separates the network into two sectors of positive (blue nodes, $V=0.5$) and negative  (yellow nodes, $V=-0.5$) voltages. (a-d) Final networks for different values of the threshold. While the sectoring of the network does not change with increasing thresholds, the altered edges decrease in number and  get increasingly clustered near the low-conductance boundary.}
    \label{fig:sector_del_1}
\end{figure}

\begin{figure}
\begin{center}
\includegraphics[scale=0.75]{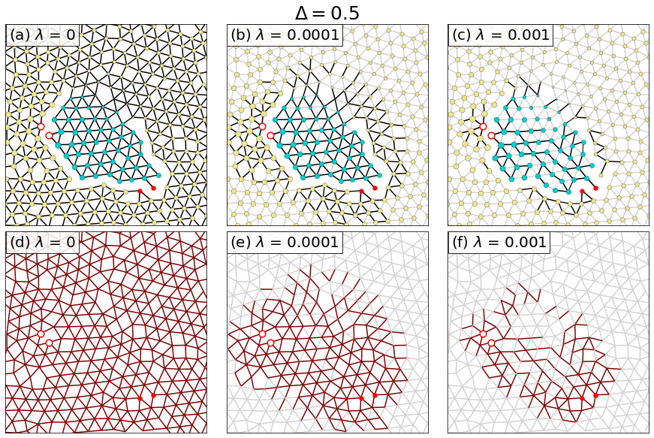}
\end{center}
\caption {Final Network with $N=1024$ nodes and $N_E=2824$  edges, after training for an edge-coupling task with $\Delta =0.5$.  Each node is depicted by a solid circle proportional in size to the absolute value of its final voltage. Each solid edge's width is proportional to its final conductance, with altered edges in black and unaltered edges in gray. The dotted edges with $k=10^{-6}$ compose the low-conductance boundary, which now only partially partitions the network into two sectors of approximately equal voltages, with positive voltages shown in blue and negative voltages shown in yellow. Again, the sectoring of the network is unchanged when the threshold is increased, while the altered edges decrease in number and become increasingly localized to the vicinity of the low-conductance boundary.}
    \label{fig:sector_del_0.5}
\end{figure}

\begin{figure}
\begin{center}
\includegraphics[scale=0.7]{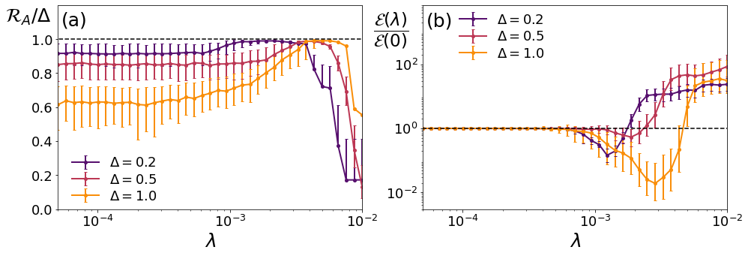}
\end{center}
\caption {Network of $N=256$ nodes and  $N_E=704$  edges, trained for two sequential edge-coupling tasks $A$ (orange) and $B$ (blue) with the same $\Delta$ value. The sources and targets of the tasks have a separation $D_A=4.0$ for task $A$, and $D_B=8.0$ for task $B$. After the full sequential training, the voltage response $\mathcal R_A$ of task $A$ (a), and the joint error $\mathcal E$ (b)  as a function of $\lambda$ shows memory improvement with thresholding for all values of $\Delta$ (color). Higher $\Delta$ values show bigger improvements in memory. Error bars represent the first and third quartiles computed from $100$ realizations.}
    \label{fig:two_tasks_del}
\end{figure}

\subsection{Thresholding Preserves the Structure of Learned Solutions}
In the main text, we showed that the learned solution of a perfect edge-coupling task ($\Delta=1$) has a distinct structure: two sectors of equal voltages separated by a low-conductance boundary, with one source and one target node in each sector, ensuring exactly equal voltage drops across the source and target nodes. 
The functionality (perfect coupling) is therefore encoded in the final sectored structure. The low-conductance boundary is not just a feature of learning dynamics but the principle structural element of the functional network, representing the largest changes made to the network during training.
In Fig.~\ref{fig:sector_del_1} we show that the structure of this learned solution, i.e. the sectoring, is not altered when a finite threshold is imposed, although fewer edges change for larger thresholds.
This is because physical learning dynamics are embedded in real space, and directed towards the sectoring of the network in edge-coupling tasks. The edges that see the largest learning signals, and are thus likely to change their conductance even with reasonably large thresholds, are those that are closest to the source, targets, and the low-conductance boundary. A finite threshold, therefore, does not alter the sectoring of the network---it only restricts learning to those edges that are more important for the distinct learned solution to develop.  
This is also true for $\Delta<1$, as Fig.~\ref{fig:sector_del_0.5} shows. In this case, the final structure is not completely separated into disconnected sectors, and yet there is an an identifiable low-conductance boundary around which the altered edges concentrate when the threshold is increased.

\subsection{Thresholding Significantly Improves Memory in Sequential Edge-coupling Tasks With High Coupling Values}
In the main text, we showed how thresholding can improve memory in sequential edge-coupling tasks with perfect coupling ($\Delta=1$). Fig.~\ref{fig:two_tasks_del} shows how memory is affected by thresholding for different values of the coupling constant $\Delta$. For these results, we consider the same sources and targets for task $A$, separated by a distance $D_A=4.0$, and randomly choose sources and targets for task $B$ with the distance between them fixed at $D_B=8.0$. For $100$ such realizations, we find that on average, the physical response of task A, $\mathcal R_A$, peaks at the desired value $\Delta$ for intermediate threshold values, with higher $\Delta$ values showing greater memory enhancement compared to the non-thresholded case ($\lambda=0$), see Fig.~\ref{fig:two_tasks_del}a.  This is because for the same distances $D_A$ and $D_B$, smaller coupling values generate smaller learning signals, such that the effect of task $B$ on the learned solution for task $A$ is small for small $\Delta$.
This facilitates good memory of task $A$ even without a threshold for small $\Delta$, whereas for big $\Delta$ values, thresholding is essential for separating the learned parameters of each task into weak or non-overlapping regions of space. Fig.~\ref{fig:two_tasks_del}b shows that in all cases, thresholding improves the joint error, regardless of $\Delta$.

\end{document}